\documentclass[prd,aps,tightline,twocolumn,nofootinbib,superscriptaddress]{revtex4}
\usepackage{graphicx}
\usepackage{amssymb}
\usepackage{amsmath}

\newcommand{\bear}{\begin{array}}  
\newcommand {\eear}{\end{array}}
\newcommand{\bea}{\begin{eqnarray}}   
\newcommand{\eea}{\end{eqnarray}}
\newcommand{\beq}{\begin{equation}}   
\newcommand{\eeq}{\end{equation}}
\newcommand{\bef}{\begin{figure}}  \newcommand 
{\eef}{\end{figure}}
\newcommand{\bec}{\begin{center}}  \newcommand 
{\eec}{\end{center}}

\begin{document}



\title{
Positron/Gamma-Ray Signatures of Dark Matter Annihilation and \\ Big-Bang Nucleosynthesis
}

\author{Junji Hisano} 
\affiliation{Institute for Cosmic Ray Research, University of Tokyo, Kashiwa 277-8582, Japan}
\affiliation{Institute for the Physics and Mathematics of the Universe,
University of Tokyo, Kashiwa 277-8568, Japan}

\author{Masahiro Kawasaki}
\affiliation{Institute for Cosmic Ray Research, University of Tokyo, Kashiwa 277-8582, Japan}
\affiliation{Institute for the Physics and Mathematics of the Universe, University of Tokyo, Kashiwa 277-8568, Japan}

\author{Kazunori Kohri}
\affiliation{Physics Department, Lancaster University, Lancaster LA1 4YB, UK}

\author{Kazunori Nakayama}
\affiliation{Institute for Cosmic Ray Research, University of Tokyo, Kashiwa 277-8582, Japan}

\date{\today}

\begin{abstract}
The positron excess observed by the PAMELA experiment may come from 
dark matter annihilation, if the annihilation cross section is large enough.
We show that the dark matter annihilation scenarios to explain the positron excess may also be compatible with the discrepancy of the cosmic lithium abundances between theory and observations.
The wino-like neutralino in the supersymmetric standard model is a good example for it.
This scenario may be confirmed by Fermi satellite experiment.
\end{abstract}

\maketitle


\section{ Introduction }
Dark matter (DM) in the Universe is one of the most striking clues to
the physics beyond the standard model (SM).  Many methods are proposed for
the direct or indirect DM detection \cite{Jungman:1995df}, and
experiments for the DM search are reaching to the
sensitivities to find an evidence of the dark matter.  Actually the HEAT \cite{Barwick:1997ig}
and PAMELA \cite{Adriani:2008zr}
experiments reported an excess of positron flux
in cosmic rays.
While  astrophysical sources, such as pulsar(s) \cite{Hooper:2008kg} or a gamma-ray burst \cite{Ioka:2008cv}, are proposed for the observed positron excess, it may be also accounted for by the high-energy positron injection from the DM annihilation \cite{Bergstrom:2008gr,Cirelli:2008pk}.

Supersymmetry (SUSY) introduces natural DM candidates as the lightest
SUSY particle (LSP).  Neutralinos in the SUSY SM are predicted to be
the LSP in many SUSY-breaking models.  The neutralino annihilation may
explain the observed positron excess. However, this generally requires
the annihilation cross section larger than expected from the thermal
relic abundance, $\langle \sigma v \rangle \simeq 3\times
10^{-26}{\rm cm^3 s^{-1}}$.

It should be noted that DM with such large annihilation
cross section significantly affects big-bang nucleosynthesis (BBN) \cite{Jedamzik:2004ip}.
(See also Ref.~\cite{Reno:1987qw} for early attempts.) 
A small fraction of the relic LSPs still
annihilates each other and injects high-energy particles into thermal
bath even after the freezeout epoch, and this may alter the abundances
of light elements significantly.  In this paper we show that the DM,
which is compatible with the positron excess, may also solve the 
discrepancy of the primordial lithium abundances between theory and
observations. The wino-like neutralino
in SUSY models is an explicit example for such DM. We notice that it may be confirmed by the gamma-ray signals from
the Galactic center by the Fermi experiment. 

\section{ Signatures of wino-like dark matter }
Wino is a superpartner for the standard-model SU(2) gauge boson. The
wino-like neutralino becomes the LSP in
anomaly-mediated SUSY-breaking models \cite{Randall:1998uk}, in which
the wino mass $(m_{\chi})$ is directly related to the gravitino
mass ($m_{3/2}$) as $m_{3/2} \sim 400 \times m_{\chi}$.  The much
heavier gravitino than the weak scale is welcome from a viewpoint of
the cosmological gravitino problem \cite{Kawasaki:2004yh}. 
The gravitino with $m_{3/2} \gtrsim 50~$TeV decays well before the BBN
begins, and the gravitino abundance after inflation is not
constrained from the observed light element abundances.

Thermal production of winos in the early Universe is not much enough to explain the
observed DM abundance, unless its mass is around 3~TeV
\cite{Hisano:2006nn}. However, even in the lighter wino cases, the
non-thermal production of winos by the gravitino decay may explain it
without spoiling the BBN.  The gravitino number-to-entropy ratio
$Y_{3/2}$ after inflation is given by $Y_{3/2}\simeq 2.3\times
10^{-14}~({T_{\rm R}}/{10^8~{\rm GeV}})$, where $T_{\rm R}$ denotes the reheating
temperature of the Universe \cite{Kawasaki:2004yh,Bolz:2000fu}.  
The current wino abundance is
almost the same as that of the gravitino since the annihilation of winos
is neglected after the gravitino decay, except for the mass range where the non-perturbative
effect significantly enhances the annihilation cross section \cite{Hisano:2006nn}.
Thus the observed DM abundance in the Universe
is explained by the non-thermal wino production if
 $T_{\rm R} \sim 10^{(9{\rm -}10)}$~GeV and $m_{\chi}\sim 100$~GeV - 2~TeV.
This value of the reheating temperature is also favored from the
thermal leptogenesis, which requires 
$T_{\rm R}\gtrsim 10^9~$GeV \cite{Fukugita:1986hr}.

Now let us discuss observational implications of wino-like DM scenario.

\subsection{ Cosmic positron flux }
The wino-like neutralinos mainly annihilate into the weak bosons, and
yields positrons, anti-protons, gamma's and neutrinos in cosmic rays, which may
give clues to the DM properties, if detected. In this paper we 
consider the positron and gamma-ray fluxes. We will comment
on the other signals later.

Energetic positrons produced by the DM annihilation lose
their energy quickly through their propagation in the Galaxy due to
synchrotron emission and inverse Compton processes with CMB photons
and star light. As a result, only positrons from the region within a few kpc can
reach to the Earth. The propagation of positrons is described by the
following diffusion equation \cite{Baltz:1998xv},
\begin{equation}
\begin{split}
        \frac{\partial}{\partial t}f(E, \vec x) 
        = &K(E)\nabla^2f(E, \vec x) \\
        &+\frac{\partial}{\partial E} [b(E)f(E, \vec x)] + Q(E,\vec x), \label{diffusion}
\end{split}
\end{equation}
where $f(E, \vec x)$ denotes the positron number density with energy $E$, 
$K(E)$ is the diffusion constant, and $b(E)$ denotes the energy loss rate.
The positron flux at the Earth (${\vec x}={\vec x_\odot}$)  is given by $\Phi^{(\rm
  DM)}_{e^+}(E,\vec x_\odot)=(c/4\pi)f(E,\vec x_\odot)$.  The source
term from the DM annihilation $Q(E,\vec x)$ is given as
\begin{equation}
        Q(E,\vec x) = \frac{1}{2} \frac{\rho^2(\vec x)}{m_{\chi}^2} \sum_f \langle \sigma v \rangle_f 
        \frac{dN^{(e^+)}_f}{dE},
\end{equation}
where $\rho(\vec x)$ is the DM mass density and 
$dN^{(e^+)}_f/dE$ is the fragmentation function of the DM annihilation
products $f$ into positrons.
We adopt the so-called M2 propagation model \cite{Delahaye:2007fr}, 
where $K(E)=0.00595~{\rm kpc^2/Myr}(E/1~{\rm GeV})^{0.55}$,
$b(E)=1\times 10^{-16}$ GeV s$^{-1}$, $L$=1~kpc ($L$ is the half-height of the diffusion cylinder)
and derive the steady state solution of Eq.~(\ref{diffusion}) 
semi-analytically \cite{Hisano:2005ec}.

The positron flux from the DM annihilation is less sensitive to the
global structure of the DM halo density profile. However, DM may
not be distributed smoothly in our Galaxy and there may be clumpy
structures in the Galactic halo.  If this is the case, the
positron flux may be enhanced \cite{Silk:1992bh}. This effect is
characterized by the boost factor, denoted by $B_F$.  Smooth
distribution corresponds to $B_F = 1$, and may reach to $\sim 5$.

\begin{figure}[ht]
 \begin{center}
   \includegraphics[width=1.0\linewidth]{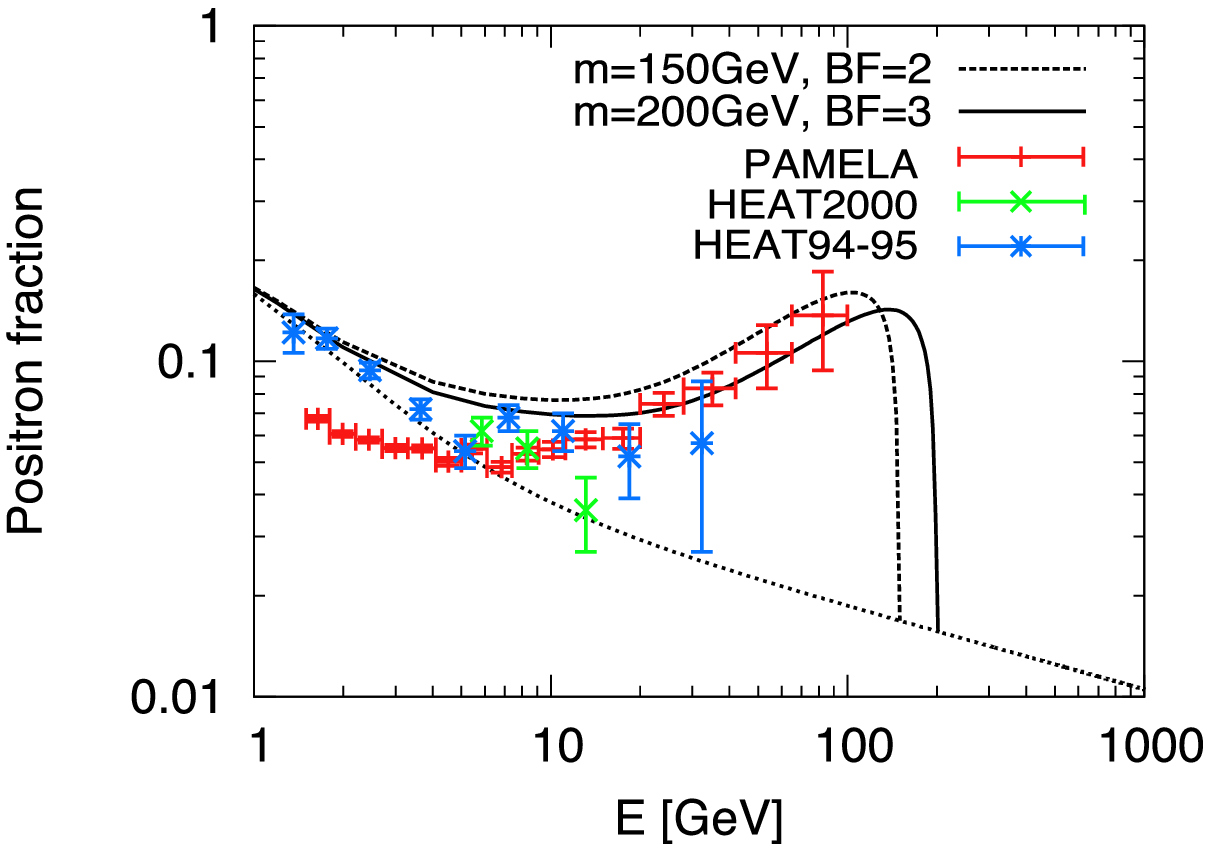} 
   \caption{ Positron fraction for $m_{\chi}=$ 150 and 200~GeV
                with boost factor 2 and 3, as a function of positron energy.
                HEAT and PAMELA results are also shown.
           }
   \label{fig:posflux}
 \end{center}
\end{figure}

Fig.~\ref{fig:posflux} shows the positron flux from the wino-like DM
annihilation using the positron fraction $R(E)$, that is the ratio of
the positron flux to sum of electrons and positrons fluxes.  The
results of the HEAT \cite{Barwick:1997ig} and PAMELA \cite{Adriani:2008zr} experiments 
are also shown.  
In the evaluation of positron fraction, we include the
background positron and electron fluxes from cosmic ray simulations
\cite{Moskalenko:1997gh}. It is found that the wino-like DM with
$m_{\chi}\sim 200$~GeV explains the PAMELA results. 
Notice that the low energy positron flux with energy less than $\lesssim 10$~GeV
is somewhat uncertain due to the solar modulation.


The ATIC balloon experiment reported an excess of the sum of the
  electron and positron fluxes, whose peak energy is around 600~GeV
  \cite{:2008zzr}.  If we believe the excess, the DM mass with
  600-1000~GeV is favored. 
  However, the ATIC excess may not be so significant if one takes
  into account large uncertainty of the data and also poor agreement with
  other experiments \cite{Torii:2008xu} in the similar energy range.
  Thus, in this paper we consider wino with
  mass around 200~GeV, since this mass range is interesting from a
  viewpoint of the cosmic lithium problem, as we will see.

\subsection{ Big-Bang Nucleosynthesis }
Even after the freezeout time of the LSPs, a small fraction of them
would still continue to annihilate each other and produce high-energy
hadrons and photons. Those emitted particles by this residual
annihilation can change the abundances of light
elements~\cite{Jedamzik:2004ip} 
such as D, T, $^{3}$He, $^{4}$He,
$^{6}$Li, $^{7}$Li and $^{7}$Be further 
after/during the BBN.

High-energy hadrons scatter off the background proton and $^{4}$He,
and induce the hadronic shower~\cite{Kawasaki:2004yh,Jedamzik:2006xz},
which produces copious neutron, D, T and $^{3}$He. The non-thermally
produced neutron and T (or $^{3}$He) scatters off the background proton
and $^{4}$He and synthesize D and $^{6}$Li, respectively. This
non-thermal neutron also induces sequential reactions to reduce
$^{7}$Be ({\it i.e.}, $^{7}$Li at a later time) through $^{7}$Be($n$,
p)$^{7}$Li(p,$^{4}$He)$^{4}$He 
(see also Ref.~\cite{Jedamzik:2004er} for the original idea).

Currently the observational $^{7}$Li abundance does not agree with the
theoretical prediction of the standard BBN when we use the
baryon-to-photon ratio, $\eta$ = $(6.225 \pm 0.170) \times 10^{-10}$,
obtained by WMAP~5-year~\cite{Dunkley:2008ie}.  Then, the theoretical
value of $^{7}$Li is much larger than the observational one even if we
adopt a relatively high value of the observational abundance,
$\log_{10}(^{7}{\rm Li/H})_{\rm obs} = -9.36 \pm
0.06$~\cite{Melendez:2004ni}. See also Ref.~\cite{Bonifacio:2006au}
for a lower value of $^{7}$Li abundance ($\log_{10}(^{7}{\rm
Li/H})_{\rm obs} = -9.90 \pm 0.06$), which is much more difficult to fit.
This situation has got worse when we use an updated reaction
rate of $^{4}$He($^{3}$He,$\gamma$)$^{7}$Be ~\cite{Li7problem}. As for
$^{6}$Li abundance, on the other hand, recent observation shows that
the theoretical value is much smaller than that of the observation,
($^{6}$Li/$^{7}$Li)$_{\rm obs}$ = 0.046 $\pm$
0.022~\cite{Asplund:2005yt}. These two discrepancies may be
collectively called ``lithium problem''. In the hadron injection
scenario, however, there is a tendency to solve the lithium problem
because it can reduce $^{7}$Li and produce $^{6}$Li as explained
above.

It should be also checked simultaneously if the abundances of the
other elements, D, $^{3}$He and $^{4}$He, meet the observational
constraints. We adopt both low and high  values of D/H, Low
(D/H)$_{\rm obs} = (2.82 \pm 0.26) \times
10^{-5}$~\cite{O'Meara:2006mj},  and   High (D/H)$_{\rm obs} =
(3.98^{+0.59}_{-0.67}) \times 10^{-5}$~\cite{Burles:1997fa}. The
observational value of the $^{4}$He mass fraction is taken to be
$Y_{p,{\rm obs}}=0.2516 \pm 0.0040$~\cite{Izotov:2007ed} with large
systematic errors~\cite{Fukugita:2006xy}. The
abundance of the $^{3}$He to D ratio is constrained by the
observational upper bound, ($^{3}$He/D)$_{\rm obs}$ = 0.83 +
0.27~\cite{2008:Geiss}.

\begin{figure}
 \begin{center}
   \includegraphics[width=1.0\linewidth]{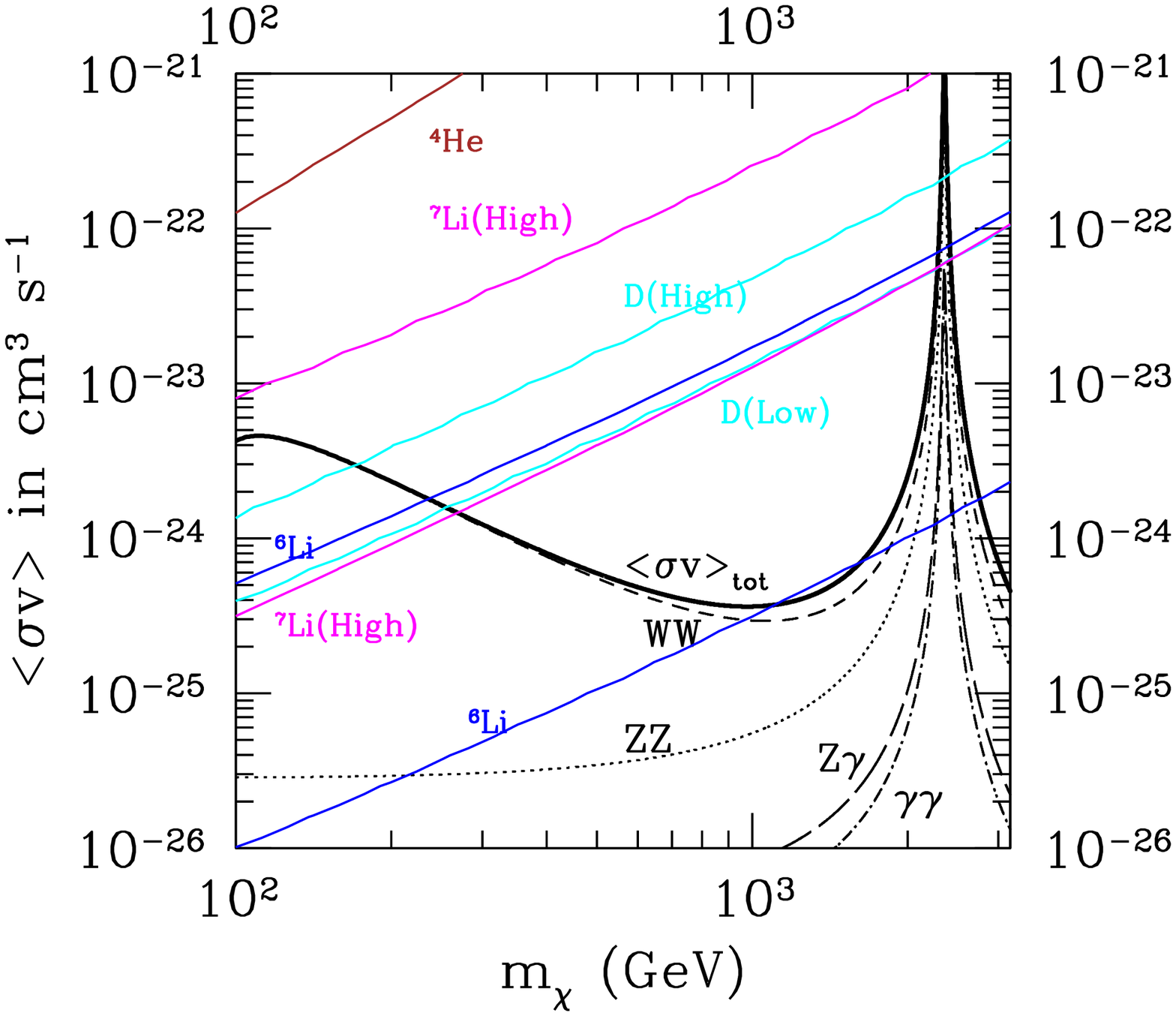}
   \vspace{-1.2cm} 
   \caption{ 
   Allowed regions at 95$\%$ C.L. from observational light element
   abundances in $m_{\chi}$--$\langle \sigma v \rangle$ plane.  The
   name of each element is written in the close vicinity of the
   line. For $^{6}$Li and $^{7}$Li,  regions sandwiched between two
   lines are allowed, respectively. Except for lithiums, each line
   means the upper bound. The total cross section of the annihilation
   and its major four modes are also plotted. The calculation is
   performed by assuming 100$\%$ $WW$ emission for simplicity. }
   \label{fig:BBNallowed}
 \end{center}
\end{figure}

The allowed region in the plane of the annihilation cross section and the
DM particle mass is shown in Fig.~\ref{fig:BBNallowed}. For
comparison, we show the wino-like neutralino annihilation cross
sections, including the non-perturbative effect on the annihilation
processes \cite{Hisano:2003ec}. Even if we
adopted the low value of D/H, it is  found that there is still an
allowed region at around $m_{\chi} \sim 250$ GeV to solve the lithium
problem, while satisfying all the constraints.

\begin{figure}
 \begin{center}
   \includegraphics[width=1.0\linewidth]{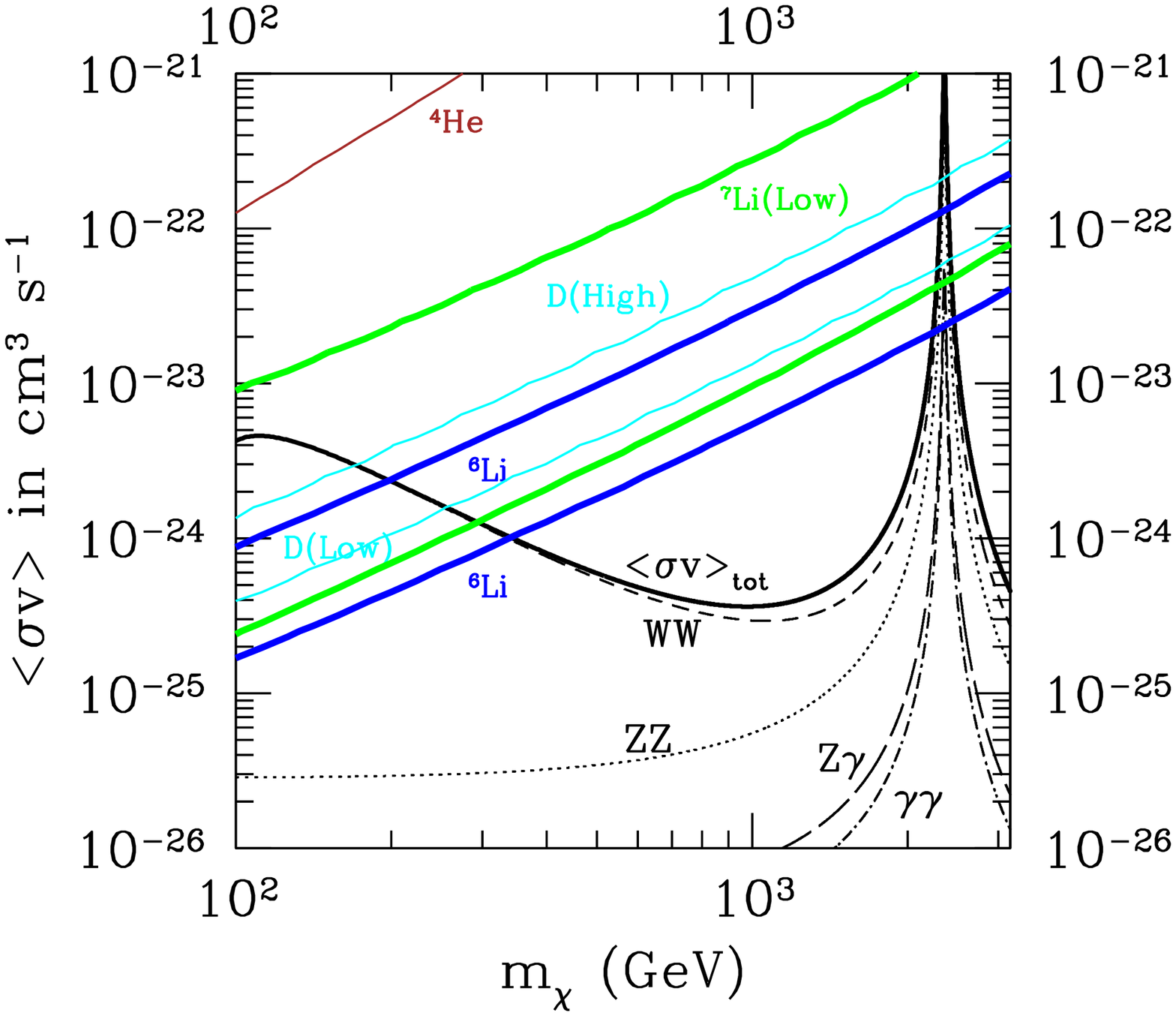}
   \vspace{-1.2cm} 
   \caption{ 
   Same as Fig.~\ref{fig:BBNallowed} except for including the depletion of
    lithium ($\Delta \log_{10}({^7{\rm Li}}/{\rm H}) 
   =0.4\Delta \log_{10}({^6{\rm Li}}/{\rm H})=0.25$).}
   \label{fig:BBNdepletion}
 \end{center}
\end{figure}

If we allow depletion of Li in stars, a larger parameter region is allowed 
as shown in Fig.~~\ref{fig:BBNdepletion}. In the figure we take the Li depletion
as $\Delta \log_{10}({^7{\rm Li}}/{\rm H})
=0.4\Delta \log_{10}({^6{\rm Li}}/{\rm H})=0.25$ which is 
implied from study of rotational mixing in stars~\cite{Pinsonneault:2001ub}.
In this case it is found that the lithium 
problem is solved even if we adopt the small value for the observed 
$^7$Li abundance for the wino mass around 150~GeV - 300~GeV.
Interestingly, the wino-like neutralino 
with this mass range can also explain the observed positron excess,
as already described.

The wino-like neutralino with mass around 2~TeV can also explain the
positron excess due to the enhancement of the cross section by the
non-perturbative effect \cite{Hisano:2005ec}. 
It is consistent with the BBN after the depletion of Li with $\Delta \log_{10}({^7{\rm Li}}/{\rm H}) \gtrsim 0.25$ is taken into account.


\subsection{ Gamma-ray flux from Galactic center }
\begin{figure}[t]
 \begin{center}
   \includegraphics[width=1.0\linewidth]{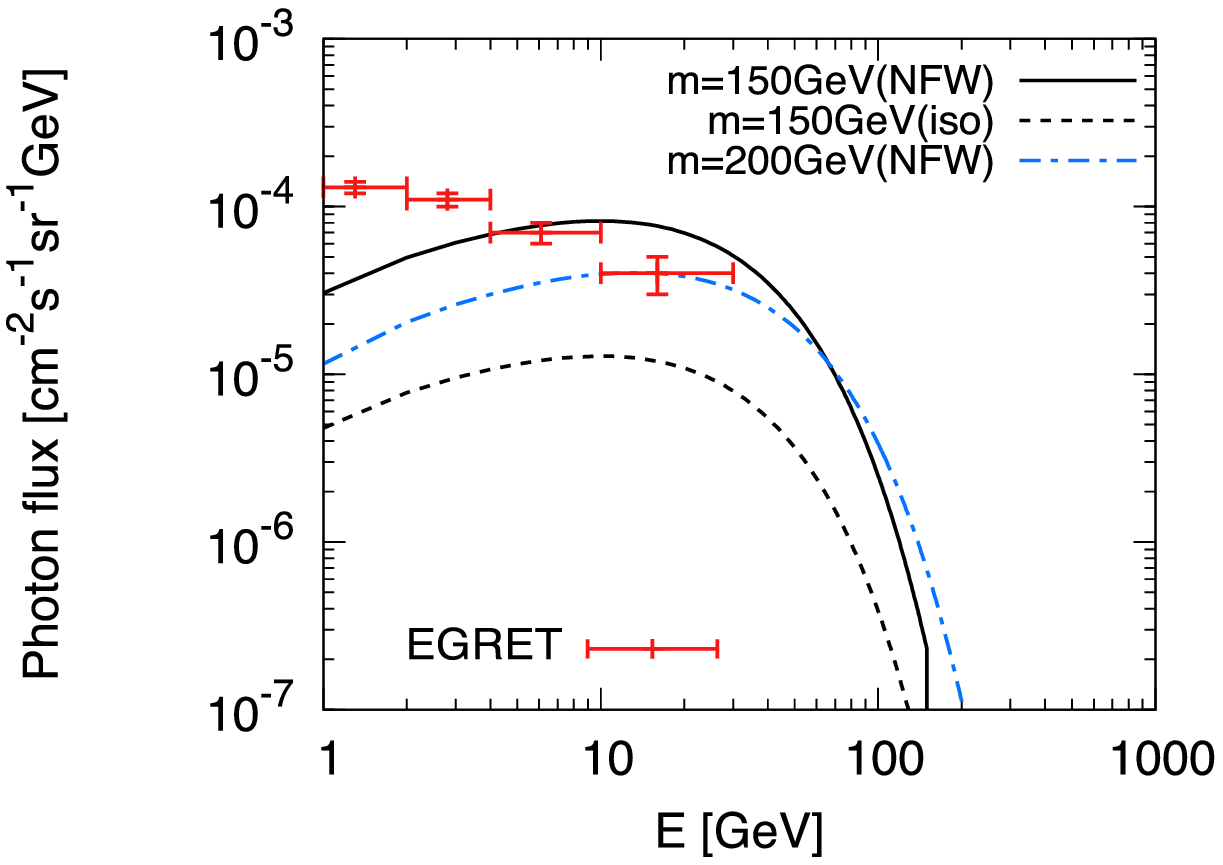} 
   \caption{ Gamma-ray flux produced by the wino-like DM annihilation
     with mass 150~GeV for both NFW and isothermal profile, 
     and 200~GeV for NFW profile from the Galactic center within
     the region $-5^\circ <l<5^\circ$ and $-2^\circ <b<2^\circ$.  
     The result of EGRET observation is also shown.}
   \label{fig:gflux}
 \end{center}
\end{figure}

DM annihilation in the Galactic halo also yields high-energy
gamma-rays.  The continuum gamma-ray flux from the neutralino
annihilation at the Galactic center is expressed as
\cite{Bergstrom:1997fj}
\begin{equation}
        \Phi_\gamma(\psi, E) = \sum_f
        \frac{\langle \sigma v \rangle _f}{8\pi m_{\chi}^{2}}\frac{dN^{(\gamma)}_f}{dE}
        \int_{\rm l.o.s.}\rho^2(l)dl(\psi),
\end{equation}
where $\psi$ is the angle from the Galactic center, $l(\psi)$ is the
distance from us along the angular direction $\psi$ and
$dN^{(\gamma)}_f/dE$ is the fragmentation function of the annihilation
products $f$ into gamma's.  The density profile $\rho$ around the
Galaxy is still unknown, and this leads to an uncertainty on the
gamma-ray flux coming from the DM annihilation at the Galactic center.
Here we consider two typical models of the DM halo: the isothermal and
Navarro-Frenk-White (NFW) profiles \cite{Navarro:1995iw}.  In
Fig.~\ref{fig:gflux} we show the gamma-ray flux from the Galactic
center  for the wino mass 150 and 200 GeV, which are favored from the 
observed positron excess and the lithium abundances.
We average the gamma-ray flux over the region of the Galactic
longitude $-5^\circ <l<5^\circ$ and latitude $-2^\circ <b<2^\circ$.
The EGRET data is also shown \cite{Hunger:1997we}.
It is seen that the gamma-ray flux is comparable to the EGRET observation depending on
the DM density profile.  It is expected that the Fermi experiment \cite{Atwood:1993zn} may
discover excess of gamma-rays and confirm the signal of DM
annihilation if the DM consists of the wino-like neutralino with mass
lighter than~300~GeV.



\section{ Conclusions and Discussion }
The positron flux excess in cosmic rays, which was first observed by
HEAT experiments and is confirmed by PAMELA experiment now, draw a
great attention of particle physicists, since it may be a striking
evidence of the dark matter.  As an example, the annihilation of
wino-like neutralino dark matter in the SUSY SM, as is realized in the
anomaly-mediated SUSY breaking models, can account for the positron
excess for the mass $m_{\chi}\sim 150$-200~GeV.
Interestingly enough, this can also solve the current discrepancy of
the primordial lithium abundances between BBN prediction and
observations.  Such models with large annihilation cross section 
also predict large gamma-ray flux from the Galactic center, which may be
observed by on-going Fermi experiments.

Some comments are in order.
The annihilation of wino-like neutralino yields $W$-bosons and they produce anti-protons,
which should be compared with observations \cite{Cirelli:2008pk}.
As opposed to the case of positron, the anti-proton flux sensitively depends on the 
choice of the diffusion zone, leading to orders of magnitude uncertainty in the 
resultant anti-proton flux \cite{Donato:2003xg}.
Within these uncertainties, the anti-proton flux from light wino DM with mass of a few hundred GeV
is consistent with observations \cite{Grajek:2008pg}.
Another constraint may come from the synchrotron radiation emitted by the electron/positrons 
from DM annihilation in the Galactic center \cite{Hooper:2007kb,Bertone:2008xr}.
However, it also suffers from large astrophysical uncertainty such as distribution of the 
Galactic magnetic field, 
which also leads to orders of magnitude uncertainty in the synchrotron flux,
and it is too early to regard the synchrotron emission
as a robust constraint on the DM annihilation model \cite{Borriello:2008gy}.
Finally, we comment on the neutrino flux coming from the DM annihilation,
which can also be constrained from the observation of Super-Kamiokande \cite{Hisano:2008ah}.
In the case of wino-like neutralino, this constraint is safely satisfied.

Although we have focused on the wino-like DM case, similar analyses
can be applied to other DM candidates.  The Higgsino-like
neutralino has about one order of magnitude smaller annihilation cross
section than that of the wino, with similar annihilation modes.  Thus,
in order to explain the positron excess by the Higgsino-like dark
matter, boost factor larger than 10 is required.  The BBN constraint is
easily satisfied in this case though the lithium problem is not
solved.  Generically, non-thermal DM production scenarios 
\cite{Moroi:1999zb} predict enhancements of the indirect signals
\cite{Profumo:2004ty,Nagai:2008se}, and such scenarios may account for the currently
observed positron excess and cosmic lithium abundances simultaneously.


The authors thank K.~Jedamzik for comments.
KN would like to thank the Japan Society for the Promotion of
Science for financial support.  This work is supported by Grant-in-Aid
for Scientific research from the Ministry of Education, Science,
Sports, and Culture (MEXT), Japan, No.14102004 (MK),
No.~20244037 (JH) and No.~2054252 (JH), and also by World Premier
International Research Center InitiativeiWPI Initiative), 
MEXT,
Japan.  KK is supported in part by PPARC grant, PP/D000394/1, EU grant
MRTN-CT-2006-035863, the European Union through the Marie Curie
Research and Training Network ``UniverseNet''.



{}

\end{document}